\documentclass[aps,prb,floatfix,twocolumn,superscriptaddress,showpacs,amsmath,amssymb]{revtex4}
\usepackage{graphicx}
\usepackage{multirow}
\usepackage{xcolor}
\usepackage{fancyhdr}

\makeatletter
\def\blfootnote{\gdef\@thefnmark{}\@footnotetext}
\makeatother

\begin{document}

\title{
Phase Stability of TiO$_2$ Polymorphs from Diffusion Quantum Monte Carlo\blfootnote{
This manuscript has been authored by UT-Battelle, LLC under Contract No. DE-AC05-00OR22725 with the U.S. Department of Energy.  The United States Government retains and the publisher, by accepting the article for publication, acknowledges that the United States Government retains a non-exclusive, paid-up, irrevocable, world-wide license to publish or reproduce the published form of this manuscript, or allow others to do so, for United States Government purposes.  The Department of Energy will provide public access to these results of federally sponsored research in accordance with the DOE Public Access Plan(http://energy.gov/downloads/doe-public-access-plan).
}
}

\author{Ye Luo}
\email{yeluo@anl.gov}
\affiliation{Argonne Leadership Computing Facility, Argonne National Laboratory, Argonne, Illinois 60439 U.S.A.}

\author{Anouar Benali}
\affiliation{Argonne Leadership Computing Facility, Argonne National Laboratory, Argonne, Illinois 60439 U.S.A.}

\author{Luke Shulenburger}
\affiliation{HEDP Theory Department, Sandia National Laboratories, Albuquerque, New Mexico 87185 U.S.A.}

\author{Jaron T. Krogel}
\affiliation{Materials Science and Technology Division, Oak Ridge National Laboratory, Oak Ridge, Tennessee 37831 U.S.A.}

\author{Olle Heinonen}
\affiliation{Material Science Division, Argonne National Laboratory, Argonne, Illinois 60439 U.S.A.}
\affiliation{Northwestern-Argonne Institute for Science and Engineering, Northwestern University, 2145 Sheridan Rd., Evanston, Illinois 60208 U.S.A.}

\author{Paul R. C. Kent}
\affiliation{Center for Nanophase Materials Sciences and Computer Science and Mathematics Division, Oak Ridge National Laboratory, Oak Ridge, Tennessee 37831 U.S.A.}

\date{\today} 


\begin{abstract}
Titanium dioxide, TiO$_2$, has multiple applications in catalysis, energy conversion and memristive devices because of its electronic structure.
Most of these applications utilize the naturally existing phases: rutile, anatase and brookite. Despite the simple form of TiO$_2$ and its wide uses, there is long-standing disagreement between theory and experiment on the energetic ordering of these phases that has never been resolved. We present the first analysis of phase stability at zero temperature using the highly accurate  many-body fixed node diffusion Quantum Monte Carlo (QMC) method. We also include the effects of temperature by calculating the Helmholtz free energy including both internal energy and vibrational contributions from density functional perturbation theory based quasi harmonic phonon calculations. Our QMC calculations find that anatase is the most stable phase at zero temperature, consistent with many previous mean-field calculations. However, at elevated temperatures, rutile becomes the most stable phase. For all finite temperatures, brookite is always the least stable phase. 

\end{abstract}

\maketitle

\section{Introduction}
Transition metal oxides are versatile compounds with a number of actual and potential applications because of their coupling between charge, spin, and lattice degrees of freedom\cite{DagottoScience2005}. Titanium dioxide, TiO$_2$, is a particular 3$d$ transition metal oxide of great interest because of its wide range of applications. 
This versatility is partly due to the many oxidation states of Ti and the different structural polymorphs of titanium oxides.   
In nature, TiO$_2$ occurs in three different structural polymorphs at ambient conditions: rutile, anatase and brookite. Rutile is the most abundant and is widely used as a white pigment, opacifier and ultraviolet radiation absorber, while anatase is the most photocatalytically active polymorph.\cite{Diebold2003,HashimotoJapJApplPhys2005} Numerous other phases have been identified or predicted, particularly at high pressure\cite{Hanaor2011,Liu2014}.

Nanocrystalline mixtures of anatase and rutile TiO$_2$ are used in dye-sensitized solar cells and water splitting for hydrogen fuel production.\cite{Hanaor2011,Gupta2011,Diebold2003} Brookite is also a good photocatalyst although it is more difficult to mine or synthesize than rutile or anatase \cite{DiPaola2013}. Because the functionality and applications of TiO$_2$ depends on the structural polymorph, accurate characterization of the phase stability of the TiO$_2$ polymorphs is critical. The most basic questions that need to be answered are which is the most stable of the three natural polymorphs, and what are the energy differences between them.

The majority of experimental studies that have addressed these questions have concluded that the bulk crystalline transition from anatase or brookite to rutile is irreversible, so that anatase and brookite are metastable.\cite{Ranade2002,Hanaor2011,Satoh2013} However, it is very challenging to accurately measure the subtle enthalpy differences between these phases. For example, the measured difference between rutile and anatase ranges from 0.158 to 2.50~mHa per TiO$_2$ formula unit (f.u.)\cite{Ranade2002,Levchenko2006}, while the JANAF tables give 2.3$\pm$0.5~mHa/f.u.\cite{JANAF}. For nanoscale and potentially hydrated samples, there are phase stability reversals due to the importance of surface contributions and relative variation in surface energies\cite{Barnard2004,Barnard2005,Barnard2008,Hummer2009,Hummer2013}.

Theoretical predictions based on density functional theory (DFT)\cite{Hohenberg1964,Kohn1965} do not give consistent results, and are often in disagreement with experimental results on the energetic ordering of the rutile, anatase, and brookite polymorphs.
Conventional DFT based on the local-density approximation (LDA) as well as many generalized gradient approximation (GGA) functionals give the result that anatase is the lowest-energy phase\cite{Muscat2002,Zhu2014}, and that brookite has lower energy than rutile\cite{Zhu2014}. This is clearly in disagreement with experiments. LDA or GGA functionals are typically inadequate to describe transition metal oxides because of the self-interaction error and strong electronic correlations in them. Specifically, LDA or GGA unphysically delocalize the 3$d$ electrons of titanium atoms. To account for electronic correlations, methods beyond LDA or GGA have been used. DFT+$U$ methods can obtain the correct energetic ordering between rutile, anatase and brookite phases by tuning the $U$-parameter\cite{Arroyo-DeDompablo2011,Curnan2015}. However, optimizing the $U$-parameter for one physical property can result in less accurate values for other physical properties. Most importantly, by its nature this approach is not predictive.
Hybrid functionals with exact Hartree-Fock exchange (EXX) often improve the accuracy of semi local DFTs. Given the generally improved thermodynamic properties of materials when calculated with hybrid functionals, it is notable that standard hybrids with typical (25\%) fractions of exact exchange do not reproduce the experimental ordering.
Indeed, a Hartree-Fock fraction of over 70\% is needed to reproduce the experimental ordering\cite{Curnan2015}.  In a recent study using a non-self-consistent random-phase approximation with EXX, it was found that rutile is more stable than anatase.\cite{Patrick2016}
The poor description of dispersion interactions afforded by DFT has also been considered to be a reason for the disagreement of the energetic ordering between LDA/GGA methods and experiments, and calculations using dispersive interactions within the DFT+D method were performed.\cite{Conesa2010,Zhu2014} However, this type of method also introduces empiricism.

Further clouding comparisons between calculations and experiments is the consideration that DFT-based calculations are usually done at 0~K. Because of the small experimental enthalpy differences between the TiO$_2$ phases, it is reasonable to expect that finite-temperature effects may play an important role and may further complicate theoretical determination of the relative stability of TiO$_2$ phases.

In order to correctly and reliably treat these small energy differences, theoretical predictions must be based on methods that are more accurate than DFT-based ones, and preferably be based on methods that provide  avenues for systematic improvements.
Quantum Monte Carlo (QMC) methods, particularly fixed-node diffusion Monte Carlo (FN-DMC)\cite{rev:qmcsolids}, offer such capability. 
FN-DMC has been successfully applied to a wide range of molecules\cite{Petruzielo2012} and solids\cite{rev:qmcsolids,luke:solids,Santana2016}.  Studies of transition metal oxides using the FN-DMC method, including Ti$_4$O$_7$\cite{BenaliTi4O7}, VO$_2$\cite{Zheng2015}, MnO\cite{Schiller2015}, FeO\cite{Kolorenc2008}, Cuprates\cite{Foyevtsova2014,Wagner2015}, NiO\cite{Mitra2015}, ZnO\cite{Santana2015,Yu2015}, have demonstrated its outstanding capability of characterizing strong electron correlation in solids.
Recent works have also demonstrated the accuracy of FN-DMC when dealing with systems dominated by dispersion interactions.\cite{Dubecky2013,Dubecky2014,Benali2014}.
Moreover, the cost to calculate the energy per atom with QMC scales as $N^3$ where $N$ is the number of electrons, and thus allows application to much larger systems than other correlated or many-body methods, such as configuration interaction (CI) and coupled cluster (CC). Because it is a Monte Carlo method, the FN-DMC algorithm can be easily parallelized and efficiently uses hundreds of thousands of processors, including both CPUs and GPUs\cite{Kim2012,Esler2012}. In brief, FN-DMC is a very promising empirical parameter-free method for simulating materials accurately and efficiently using supercomputers.

In this work, we first study the phase stability of rutile, anatase and brookite polymorphs at zero temperature using the highly accurate FN-DMC. We determine the energetic ordering after considering all sources of systematic errors, which can potentially contaminate the high accuracy required in order to unambiguously determine the subtle energy difference between these different phases. We find anatase to be the most stable polymorph at 0~K, but we find the energy difference between anatase and rutile only to be 1.5~mHa per formula unit (f.u.) with a statistical uncertainty of 0.1~mHa/f.u., while rutile and brookite have the same energy within statistical errors in the absence of zero point motion.  
In order to account for finite temperature and to determine the 
phase stability of the polymorphs at finite temperatures, we also study the phonon contributions to the phase stability of all the three polymorphs at finite temperatures using density functional perturbation theory (DFPT)\cite{Baroni2001}. Temperature has a profound effect on the phase stability, producing a different ordering than at zero temperature.  Taking phonon contributions into account, we find that the free energy of rutile becomes lower than that of anatase at 650$\pm 150$~K. This is consistent with the experimental observations that rutile can easily be synthesized from anatase by keeping it at above 870~K for a few minutes up to a day or at 663~K after one week.\cite{Hanaor2011} We also find that the free energy of brookite is higher than that of rutile and anatase at all finite temperatures.

\section{Computational methods}
\label{sec:method}
\subsection{Quantum Monte Carlo methods}
The QMC calculations reported here are carried out within the variational Monte Carlo (VMC) and FN-DMC frameworks implemented in the QMCPACK code,\cite{Kim2012,Esler2012}
with a Slater-Jastrow type trial wave function.
The single-particle orbitals used to construct the Slater determinants are extracted from LDA+$U$\cite{qe:lda+u} calculations obtained from the Quantum ESPRESSO\cite{qe:main} electronic structure code. 
The plane-wave cutoff in DFT/LDA calculations was 150 Ha which is much higher than conventional DFT calculations because of the hard pseudopotentials we use here. Total energies were converged to 0.01~mHa/f.u.\ with respect to this cutoff. 
The Monkhorst-Pack meshes of $k$-points are $8 \times 8  \times 8$, $12 \times 12 \times 6$ and $6 \times 8 \times 8$
 for rutile, anatase and brookite primitive cells respectively. 
Total energies were converged to 0.0005~mHa/f.u.\ with these k-point samplings.
The Hubbard $U$ parameter is chosen variationally by FN-DMC\cite{BenaliTi4O7} (see Sec.~\ref{sec:LDA+U}), which provides
an un-biased way of determining this parameter. Note that this method does not guarantee that DFT+$U$ calculations with this value of $U$ will yield any better agreement with experiments than other methods\cite{qe:lda+u,Campo2010}. This method instead is a limited way to optimize the nodal surface of the FN-DMC wave function. Within QMCPACK, the orbitals are evaluated on a real space mesh using b-splines to achieve a constant-time per orbital evaluation of the orbitals independent of the basis set size.
The Jastrow part of the trial wave function contains both one- and two-body Jastrow factors with a total of 40 parameters optimized by energy minimization \cite{Umrigar:linear} within VMC. 
The one-body Jastrow has a cutoff radius as large as the Wigner-Seitz (WS) radius of a 16~f.u.\ cell, listed in Table~\ref{tab:cellspec}. The two-body Jastrow cutoff is the same as the WS radius of the supercell.
To test the sensitivity of the DMC energy to the Jastrow factor, we also used a three-body Jastrow factor to capture additional ion-electron-electron correlations; this calculation was however limited to 36~f.u.\ supercells because of its higher computational expense.
The three-body Jastrow factor lowers the final DMC energies, but the energy difference between rutile and anatase remains consistent with the calculation with only one and two body Jastrow factors, 2.2(3) vs 2.5(3)~mHa, see Table~\ref{tab:J3}. All the optimized Jastrow factors are provided in the Supplemental Material.

\begin{table}
\centering
\footnotesize
\begin{tabular}{c|c|ccc}
\hline
\multirow{2}{*}{Polymorph} & \multirow{2}{*}{Lattice parameters (\AA)} & \multicolumn{3}{c}{WS radii (Bohr)}\\
\cline{3-5}
& & S16 & S32/36 & S72 \\
\hline
rutile\cite{Isaak1998}(RT) & a,c=4.5938,2.9586 & 7.0780 & 9.7057 & 12.2768 \\
anatase\cite{arlt2000}(298K) & a,c=3.78512,9.51185 & 7.1528 & 8.9874 & 12.0303 \\
brookite\cite{Baur1961}(RT) & a,b,c=9.184,5.447,5.145 & 7.0796 & 9.7226 & 11.3836 \\
\hline
\multicolumn{5}{l}{RT: Room temperature.}
\end{tabular}
\caption{Lattice parameters of the unit cells and Wigner-Seitz radii of supercells for rutile, anatase and brookite. Rutile, anatase, and brookite has unit cells of 2, 4, and 8 f.u., respectively.
The last three columns show the Wigner-Seitz radius for each computational cell containing 16, 32 (brookite), 36 (rutile and anatase), and 72 formula units.}
\label{tab:cellspec}
\end{table}

\begin{table}
\centering
\begin{tabular}{c|c|c|c}
\hline
\multirow{2}{*}{$\psi_T$} & \multicolumn{2}{c|}{E(TiO$_2$) (Ha)} & \multirow{2}{*}{$\Delta E$ (Ha)}\\
\cline{2-3}
& rutile & anatase & \\
\hline
SD+J12 & -90.6192(2) & -90.6217(2) & -0.0025(3) \\
SD+J123 & -90.6220(2) & -90.6242(2) & -0.0022(3) \\ 
\hline
\end{tabular}
\caption{FN-DMC energies of rutile and anatase (TiO$_2$)$_{36}$ with (SD+J123) and without (SD+J12) a three-body Jastrow factor for
a 36-formula unit unit cell; a time step 0.004 Ha$^{-1}$ is used in these calculations.}
\label{tab:J3}
\end{table}

In our calculations, the core electrons, (1s,2s,2p) of Ti and (1s) of O, respectively, are removed by using scalar relativistic norm-conserving pseudopotentials(PPs)
generated by DFT-LDA atomic calculations and customized for QMC by using very small cutoff radii for more accurate core description (labeled as PP-QMC).\cite{Krogel2016,Root2015} These pseudopotentials are provided together with benchmark results\cite{Krogel2016,Root2015} in the Supplemental Material. To assess the pseudopotential bias on phase stability, we have calculated the energy difference between anatase and rutile in 16~f.u.\ supercells using both high-fidelity Burkatzki, Filippi, Dolg (BFD) Hartree-Fock PPs\cite{Burkatzki2007,Burkatzki2008} and our state-of-art LDA PPs.
Both set of PPs give the same energy difference between the phases with our target accuracy. This indicates that the difference between these potentials is not biasing the results. Therefore our findings about phase stability are reliable.

\begin{table}[t]
\centering
\begin{tabular}{c|c|c|c}
\hline
PP & Anatase (Ha) & Rutile (Ha) & diff.\ (mHa) \\
\hline
Ours & -90.63490(24) &  -90.63243(12) & 2.47(27) \\
BFD  & -90.7077(1)  &  -90.7052(1)  & 2.50(13) \\
\hline 
\end{tabular}
\caption{DMC energy values of anatase and rutile with the Chiesa correction calculated in 16~f.u.\ supercell and time step 0.01~Ha$^{-1}$ using BFD and our PPs. The error bars of calculations with BFD PPs are smaller due to more statistics collected. Though the absolute energy values are different, the energy difference is consistent within statistical error below our accuracy criteria 0.37~mHa.}
\label{tab:AnRu}
\end{table}

To allow the use of non-local pseudopotentials, the locality approximation\cite{Mitas:LA} is often applied. However, it results in non-variational energies and can result in dynamical instabilities in the walker population.
Therefore, we instead use T-moves\cite{Casula:Tmove}, which are devised to directly treat the negative sign from non-local operators by an effective Hamiltonian, restoring the variational nature of the method and regaining a stable walker population. 

\begin{figure}[t]
\centering
\includegraphics[width=\columnwidth]{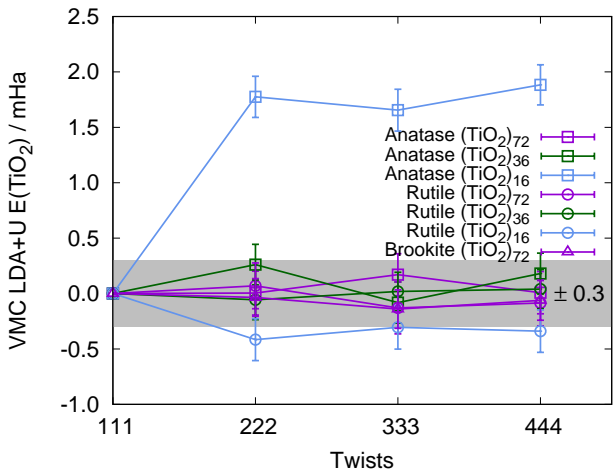}
\caption{The VMC energy difference per formula unit between the multiple twist grids and the Gamma point. The energy value of (TiO$_2$)$_{16}$ anatase converges using a $2\times2\times2$ twist grid; all the other energy values of anatase, rutile and brookite calculated at the Gamma point are well converged within 0.3~mHa (shaded gray region).}
\label{fig:twists}
\end{figure}

The accuracy of DMC calculations applied to periodic solids are affected by two types of finite size errors. These arise whenever the thermodynamic-limit behavior of a solid is approximated from calculations on a finite size simulation cell with periodic boundary conditions. The first type are errors from
one-body finite-size effects. These can be mitigated using twist-averaged boundary conditions\cite{Ceperley:twists}.
By comparing the ground state VMC energies averaged on different twist grids, 
we find that the energy evaluated with only the gamma point has already converged to our required accuracy of 0.37~mHa per f.u.\ 
for all of the three gapped TiO$_2$ polymorphs when using medium and large supercells, see Fig.~\ref{fig:twists}. In the simulations with the smallest 16 formula unit supercell, the Gamma point is sufficient for rutile but a $2 \times 2 \times 2$ twist grid is needed for anatase. All of the following FN-DMC calculations obey this choice of twists.

The other type of finite-size errors in DMC calculations are from two-body effects that arise from artificial periodic
images of an electron's exchange-correlation hole. This error is more difficult to control than the one-body ones.
Many different schemes have been devised to control these errors, including the modified periodic Coulomb (MPC)\cite{Fraser1996,Kent1999,Drummond2008} and Chiesa\cite{Chiesa2006} corrections to mitigate the two-body potential or kinetic energy errors. The former modifies the Coulomb potential while the latter directly analyzes the long range Jastrow factor and electronic structure factor to determine corrections.
Unless otherwise specified, the latter one is used in all our calculations for the reason explained in Fig.~\ref{fig:finite_diff}. We perform calculations with three supercell sizes and extrapolate values to infinite size. Finite size extrapolation is essential, regardless of the finite size correction scheme used, and at least three cell sizes are needed to judge the accuracy of the extrapolation. The shapes of the supercells are generated in an effort to maximize simulation cell radius for a given number of atoms in the supercell in order to further reduce two-body finite-size effects. The tiling matrices used to construct the different supercells are given in the Supplemental Material.

We perform FN-DMC calculations of rutile, anatase and brookite solids each in three supercell sizes specified in Table~\ref{tab:cellspec}. 
The largest supercell, 72 formula units, contains 216 atoms and 1728 electrons.
The lattice constants of the primitive cells were taken from experiments.\cite{Isaak1998,arlt2000,Baur1961} 
For consistency with our later phonon calculations, the coordinates of all the atoms in the unit cell are relaxed with DFT/LDA. The residual forces are smaller than 0.1~mHa/a.u.\ on each atom after the relaxation.
For rutile and anatase, the Wyckoff positions, listed in Table~\ref{tab:atomic_positions}, given by DFT/LDA, DFT/PBE\cite{Zhu2014}, and the experiment\cite{Burdett1987} differ by less than 0.001.
The Wyckoff positions of brookite given by DFT/LDA, DFT/PBE\cite{Zhu2014}, and the experiment\cite{Wyckoff1963} have slightly larger differences, but are still within 0.005. 
The DFT/LDA energy differences between the experimentally determined structures and the ones relaxed with LDA are 0.005, 0.03, and 0.1~mHa/f.u.\ for rutile, anatase, and brookite, respectively. These differences are substantially smaller than our target accuracy 0.37~mHa, indicating that use of LDA relaxed structures will not significantly influence the phase stability.

\begin{table}
\centering
\footnotesize
\begin{tabular}{c|c}
\hline
Polymorph & Wyckoff positions \\
\hline
rutile & Ti 2a(0,0,0) \\
($P4_2/mnm$)& O 4f(x,x,0): x=0.3051$^a$, 0.3052$^b$, 0.3047(1)$^c$ \\
\hline
anatase & Ti 4a(0,0,0) \\
($I4_1/amd$)& O 8e(0,0,z): z=0.2074$^a$, 0.2068$^b$, 0.2081(1)$^c$ \\
\hline
brookite & Ti 8c(x,y,z): x=0.1289$^a$, 0.1290$^b$, 0.1290$^d$ \\
($Pbca$)& y=0.0951$^a$, 0.0906$^b$, 0.0972$^d$\\
& z=0.8624$^a$, 0.8623$^b$, 0.8629$^d$\\
& O1 8c(x,y,z): x=0.0116$^a$, 0.0100$^b$, 0.0101$^d$ \\
& y=0.1469$^a$, 0.1483$^b$, 0.1486$^d$\\
& z=0.1826$^a$, 0.1833$^b$, 0.1824$^d$\\
& O2 8c(x,y,z): x=0.2294$^a$, 0.2298$^b$, 0.2304$^d$ \\
& y=0.1073$^a$, 0.1081$^b$, 0.1130$^d$\\
& z=0.5354$^a$, 0.5365$^b$, 0.5371$^d$\\
\hline
\end{tabular}
\caption{Atomic positions of rutile, anatase, and brookite. a) relaxed with LDA, a) relaxed with PBE\cite{Zhu2014}, c) experiment\cite{Burdett1987}, d) experiment\cite{Wyckoff1963}.}
\label{tab:atomic_positions}
\end{table}

\subsection{Trial wave function optimization}
\label{sec:LDA+U}

When studying fermionic systems, the fixed-node approximation is applied in order to ensure the antisymmetry of the trial wave function in the DMC calculations. The FN-DMC total energy is variational with respect to the quality of the nodes of the trial wave function.
With the Slater-Jastrow trial wave functions used here, the trial wave function consists of a single Slater determinant multiplied by Jastrow factors. 
The Jastrow factors are never zero and therefore do not alter the nodal structure.
The nodes are controlled by the single particle orbitals within the Slater determinant. Currently, the complete optimization of these orbitals within Quantum Monte Carlo is not practical for our simulation sizes. Instead, we use orbitals extracted from LDA+$U$ calculations. The best set of orbitals gives the lowest DMC energy. 

Within DFT studies using the DFT+$U$ framework, it is always a big challenge to justify the right value of parameter $U$ as different methods may yield very distinct values even when the same system is studied.
A value of $U$ can be computed in a self-consistent manner through a linear response method \cite{qe:lda+u,Campo2010}, 
 or by using an iterative scheme combining DFT+$U$ and GW calculations to minimize the quasi-particle corrections \cite{Patrick2012}. 
Another route is to empirically select the $U$ value so as to reproduce  experimental results of a physical property, such as band gaps\cite{Vu2012}, or reaction energy\cite{Li2006}.
Because of the complexity (and ambiguity) of choosing a value for $U$, a wide range of $U$ from 2.5~eV to 7.5~eV has been reported in previous studies on TiO$_2$ polymorphs.\cite{Zhu2014,Arroyo-DeDompablo2011,Vu2012,Patrick2012}

In contrast, within the FN-DMC framework $U$ is a variational parameter in the total energy. When the T-move scheme is used for their evaluation\cite{Casula:Tmove}, the total energy evaluated by FN-DMC is fully variational even when pseudopotentials are used. A better nodal surface described by a more accurate trial wave function yields a lower FN-DMC energy. Therefore, the  value of $U$ which minimizes the FN-DMC energy gives the best orbitals for constructing the trial wave function. This practice has been applied in many previous studies on transition metal oxides, e.g.\ Refs.~\onlinecite{Foyevtsova2014, Santana2015, BenaliTi4O7}.
Other works using orbitals obtained from hybrid functionals show that varying the weight of the exact exchange has the similar effect as varying $U$ in DFT+$U$ for optimizing the nodal structure, e.g.\ Refs.~\onlinecite{Kolorenc2008, Zheng2015, Wagner2015, Schiller2015, Yu2015}. This quality of wave function is sufficient to reproduce phase transitions in FeO\cite{Kolorenc2008}, the metal-insulator transition in VO$_2$\cite{Zheng2015}, magnetic properties of cuprates\cite{Foyevtsova2014,Wagner2015}, the phase diagram of MnO\cite{Schiller2015},  the properties of ZnSe and ZnO\cite{Yu2015,Santana2015}, and group IIA and IIIB binary oxides\cite{Santana2016}. We therefore expect this quality of wave function to be sufficient for the study of TiO$_2$.

\begin{figure}[t]
\centering
\includegraphics[width=\columnwidth]{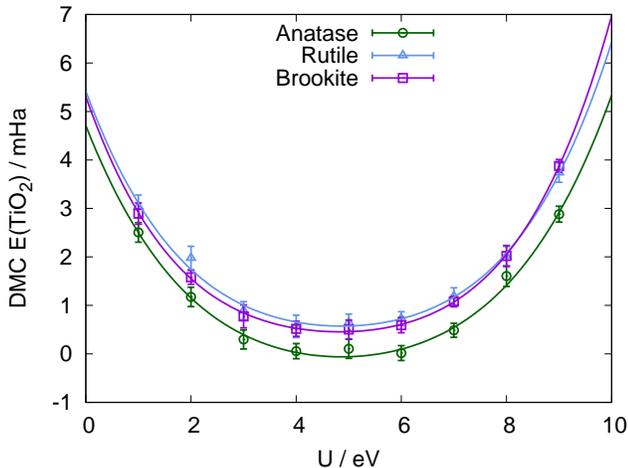}
\caption{FN-DMC energy per TiO$_2$ formula unit as a function of the $U$ parameter with a time step 0.01~Ha$^{-1}$.
The calculations of rutile and anatase are performed with 36 f.u.\ supercells, and brookite has a different supercell with 32 formula units. The curves show a fit using a quartic polynomial function. The values of $U$ corresponding to the energy minimum of rutile, anatase and brookite are 4.86(19), 4.86(20) and 4.83(6)~eV respectively. The energy values are shifted by -90.6369~Ha to place the energy minimum of anatase at zero energy.}
\label{fig:getU}
\end{figure}

In this work, for each of the three polymorphs, we constructed a set of trial wave functions varying the value of $U$ from 1.0~eV to 9.0~eV in DMC calculations using cells of 32 or 36 f.u.\ (Fig.~\ref{fig:getU}). The energy minimum is reached at $U$ equal to 4.86(19), 4.86(20) and 4.83(6)~eV for rutile, anatase and brookite, respectively. Because all three values are consistent within error bars we use $U=4.86$ eV in all of the following calculations. We note that this data already clearly shows anatase as lowest in energy.

\section{Results}
\subsection{Time step error}
\begin{figure}[t]
\centering
\includegraphics[width=\columnwidth]{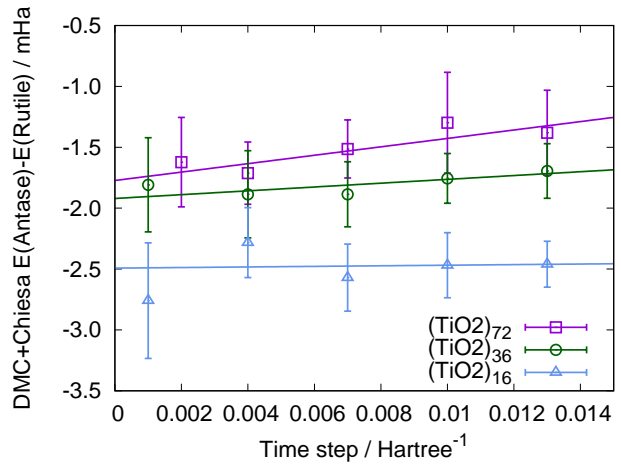}
\caption{FN-DMC energy difference per formula unit between anatase and rutile as a function of the DMC time step. The extrapolated energy differences at zero time step are $-1.77(9)$, $-1.92(7)$ and $-2.49(16)$~mHa for supercell sizes 72, 36 and 16 respectively.}
\label{fig:tstep}
\end{figure}

As a Green's function projector method for solving the Schr\"odinger equation in imaginary time,
the DMC algorithm is accurate only in the limit of small time steps $\tau$. 
However, the computational effort required to achieve a given error bar scales with $1/\tau$ and the straightforward use of infinitesimally small time steps is not possible.
In practice, three or more sufficiently small time steps $\{\tau_i\}$ are used and the ground-state energy is obtained by  extrapolating to the limit $\tau = 0$.
Because of the subtle energy difference between rutile and anatase, the most valuable calculations are performed on the large supercells 
in order to reduce all sources of finite size error while a  statistical error below 0.37~mHa per formula unit is required.
Hence, using very small time steps (reaching the linear regime) becomes extremely expensive in terms of computational cost. 
Instead, we monitor the convergence of the energy differences between the phases as the time step is reduced. 
We calculate the FN-DMC total energy difference between rutile and anatase in the three supercell sizes with $\tau = 0.004, 0.007, 0.010, 0.013$~Ha$^{-1}$,
 as well as $0.001$ for the small and medium sizes and $0.002$ for the large size; see data points and fits to linear functions in Fig.~\ref{fig:tstep}. This figure shows that the energy differences in all the three sizes are very robust as $\tau$ decreases and the time step error with our chosen time steps does not affect the energy ordering: anatase is always lowest in energy and not sensitive to the cell size or time step used. 

\subsection{Finite-size effects}
\begin{figure}[t]
\centering
\includegraphics[width=\columnwidth]{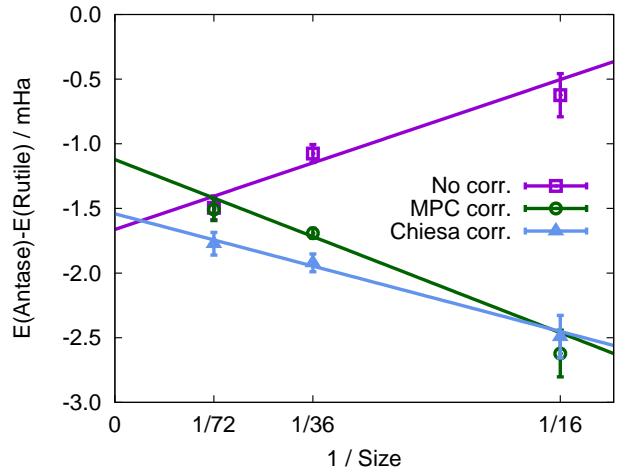}
\caption{FN-DMC energy difference per formula unit between anatase and rutile as a function of supercell size. The zero time step data points are fitted linearly and extrapolated to infinite supercell size. MPC and Chiesa corrections are both added individually to correct the finite size error. The extrapolated energy values with Chiesa and without correction are consistent within error bars but the MPC one is higher. For this reason, we choose Chiesa correction as our scheme used in all the calculations. The energy difference $E_{\rm Anatase}-E_{\rm Rutile}$ extrapolated to both zero time-step and infinite supercell is $-1.5(1)$~mHa.}
\label{fig:finite_diff}
\end{figure}

FN-DMC is a real space method and therefore suffers from finite-size effects when applied within periodic boundary conditions. We therefore perform calculations on a series of different sized supercells and extrapolate to infinite cell size. Fig.~\ref{fig:finite_diff} shows the FN-DMC energy difference between rutile and anatase calculated using the uncorrected Coulomb interaction, MPC and Chiesa schemes for multiple cell sizes. In our largest calculations of 72 formula units the different finite-size correction schemes agree to within 0.4~mHa. Calculations on smaller cell sizes are more sensitive to the finite-size correction schemes. However, the sign of the energy difference is insensitive to cell size and to the finite size correction scheme.

To reduce the computational cost of studying brookite, we performed the calculation of brookite solid only at $\tau=0.004$~Ha$^{-1}$ in a 72 formula unit supercell because the largest supercells have energies very close to the value obtained when extrapolated to infinite cell size. The total energy values of rutile, anatase and brookite for final comparison are reported in Table~\ref{tab:final}.
Anatase has the lowest energy while rutile and brookite have the same energy within statistical errors. Therefore, anatase is the most stable phase at zero temperature based on our FN-DMC calculations. This is the same qualitative result as is obtained using conventional DFT with LDA/GGA functionals and the ``incorrect'' order as compared to many experimental reports.
FN-DMC predicts that rutile and brookite are very competitive in stability at zero temperature while DFT favors brookite\cite{Zhu2014}.
Additional discussion is given in sec.~\ref{sec:lattice_dynamics}, after considering the role of lattice dynamics at zero and elevated temperatures.

\begin{table}[t]
\centering
\begin{tabular}{c|c|c|c}
\hline
\multirow{2}{*}{Energy} & \multicolumn{3}{c}{E(TiO$_2$)/Ha} \\
\cline{2-4}
& rutile & anatase & brookite\\
\hline
LDA (TiO2) & -92.10711 & -92.10852 & -92.10826 \\
\hline
DMC (TiO$_2$)$_{72}^*$ & -90.6285(2) & -90.6302(2) & -90.6289(1) \\ 
\hline
DMC (TiO$_2$)$_{72}$ & -90.6300(2) & -90.6315(2) & -90.6300(1) \\ 
ZPE & 0.00766 & 0.00798 & 0.00800 \\
\hline
Total & -90.6223(2) & -90.6235(2) & -90.6220(1) \\
\hline
\end{tabular}
\caption{FN-DMC energies of rutile, anatase and brookite (TiO$_2$)$_{72}$ calculated with a time step 0.004~Ha$^{-1}$, and zero point energy (ZPE) calculated using DFPT. Energy values with (labeled with $^*$) and without Chiesa correction are provided. Because the energy difference between anatase and rutile without correction are consistent with the extrapolated value obtained in Fig.~\ref{fig:finite_diff}, we use all the three DMC energy values without Chiesa correction for the internal energy correction in the phonon study. The energy values given by DFT with PP-GBRV are also listed here.}
\label{tab:final}
\end{table}

\subsection{Lattice dynamics and finite-temperature enthalpies}
\label{sec:lattice_dynamics}
While our DMC calculations help to 
establish the energy ordering of the rutile and anatase phases at 0~K (neglecting zero point energy), our results are in disagreement with experimental observations at finite temperature\cite{Ranade2002,Hanaor2011,Levchenko2006,Satoh2013} in which rutile is the most stable phase. We therefore investigate the role of
lattice vibrations at finite temperature on the phase stability to see if the ordering is influenced more by the entropy  than the electronic contribution to the free energy at finite temperatures. The use of QMC for studying lattice dynamics would be desirable, but such calculations are not yet possible because of the lack of developed algorithms. We therefore utilize DFT for the finite temperature study.

We calculate the Helmholtz free energy $H(V,T)$ at a temperature
 $T$ and volume $V$ within the quasi-harmonic approximation (QHA), in which
\begin{eqnarray*}
   H(V,T)&=&U(V)+\frac{1}{2}\sum_{\vec{q},j}h\omega_j(V,\vec{q}) \\
       && +k_B T \sum_{\vec{q},j}\ln \Big[1 - \exp\big(- h\omega_j(V,\vec{q})/k_B T\big)\Big],
\end{eqnarray*}
where the first, second and third terms are, respectively, the internal, zero-point and thermal contributions. In our calculations, the internal energy is calculated as
\begin{equation*}
     U(V)= U^{\rm DFT}(V)+U^{\rm QMC}(V_{\rm exp})-U^{\rm DFT}(V_{\rm exp}),
\end{equation*}
where $U^{\rm DFT}(V)$ is the DFT internal energy, amended by adding the energy difference between FN-DMC and DFT on the experimental structure.  In general this energy correction should be volume dependent, but this approximation is accurate for the very small volume range considered here.
The zero-point and thermal contribution terms are obtained using density functional perturbation theory (DFPT) as implemented in the phonon code of the Quantum ESPRESSO package\cite{qe:main}.
The accuracy of DFT is sufficient for computing the inter-atomic force constants (IFC) used by phonon calculations because the phonon dispersion relations of rutile calculated by DFT and measured by inelastic neutron scattering are in a good agreement\cite{Sikora2005}.

For these calculations we utilize GBRV ultra-soft LDA pseudopotentials\cite{GBRV} and 40~Ha kinetic energy cutoff for lower computational cost without compromising accuracy. Calculations with PP-GBRV and our much harder PP-QMC (generated by DFT/LDA) indicate the consistency of pseudopotentials in DFT and QMC calculations by showing no difference in the phonon dispersion relation and zero-point energy. DFPT calculations with Helmholtz free energy converged to 0.05~mHa/f.u.\ are obtained with  $k$- and $q$-point meshes of $8\times 8\times 8$ and $6\times 6\times 6$ for rutile, $12\times 12\times 6$ and $4\times 4\times 2$ for anatase, and $4\times 6\times 6$ and $2\times 4\times 4$ for brookite.
The final phonon density of states is recalculated from the IFC on a much denser $q$-mesh, $13\times 13\times 13$.
To apply the QHA, phonon calculations are performed on eight volumes for each solid with the shape of cell and atomic positions fully relaxed. Note that for rutile, given the instability caused by the soft modes in large expansion\cite{Mitev2010}, free energy values at large expansion are extrapolated.
The fact that these DFT-predicted soft phonon modes have been confirmed by inelastic X-ray scattering\cite{Wehinger2016} strengthens the credibility of DFT phonon calculations in these materials.

\begin{figure}[t]
\centering
\includegraphics[width=\columnwidth]{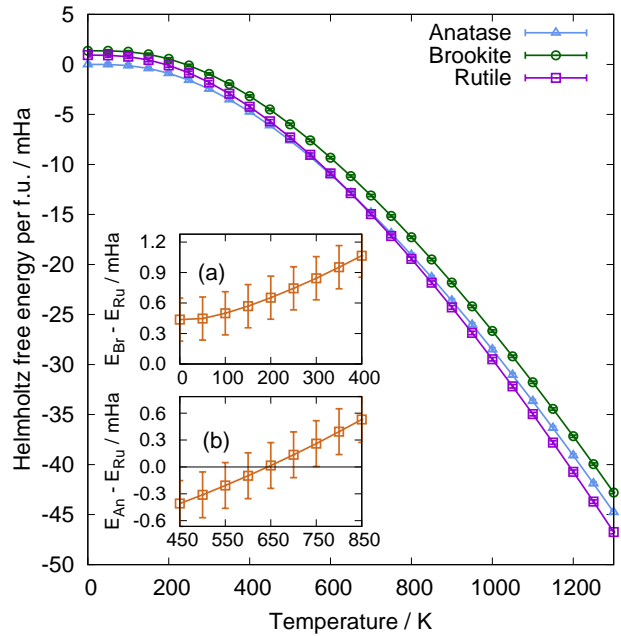}
\caption{The Helmholtz free energy of rutile, anatase and brookite as a function of temperature. All the values are shifted by -90.6241~Ha, the Helmholtz free energy of anatase at 0~K. The energy differences between brookite and rutile at 0--400~K, and between anatase and rutile at 450--850~K are provided in the insets (a) and (b). The energy of brookite is always larger than that of the other two solids, while the energy of rutile becomes lower than that of anatase at 650$\pm$150~K. The error bars indicate the statistical uncertainty due to the QMC data used for the 0~K energy differences.}
\label{fig:lattice_dynamics}
\end{figure}

The zero point energy (ZPE) values of rutile, anatase and brookite are listed in Table~\ref{tab:final}, and the Helmholtz free energy as a function of temperature are shown in Fig.~\ref{fig:lattice_dynamics}.
At zero temperature, anatase remains the most stable phase even though the larger ZPE of anatase reduces the energy difference between anatase and rutile, but it is not enough to change the energetic ordering qualitatively. Because of a larger ZPE, brookite becomes less stable than rutile, which is consistent with the experimental findings.
As the temperature is increased, the Helmholtz energy of rutile decreases much faster than that of anatase, and above $650\pm 150$~K, rutile becomes the most stable phase. This phase transition is consistent with many experimental observations\cite{Hanaor2011,Satoh2013} that rutile can be very easily synthesized by heating anatase above about 870~K. 
It was also reported that at 663~K, rutile became detectable in anatase powder after one week.\cite{Hanaor2011} 
In all temperatures up to the melting point, brookite remains less stable than rutile and anatase.

We would like to emphasize that the QMC correction for the internal energy is crucial to improve the theoretical prediction of the phase stability between rutile and brookite,
 although it does not change the relative stability of anatase and rutile compared to pure DFPT. 
This experience clearly shows that a specific density functional  may work well in a few cases but is not guaranteed to work all the time. Thus methods like QMC offering higher reliability and predictability are desired. 

\section{Conclusions and Outlook}

We have studied the phase stability of rutile, anatase and brookite polymorphs of TiO$_2$ at 0~K using highly accurate FN-DMC calculations. We also calculated the Helmholtz free energy of the polymorphs at finite temperature by including a DFPT based quasi-harmonic treatment of the phonon contributions. Our results show that at zero temperature, anatase is the most stable phase, while brookite is less stable than rutile when zero-point energy is included. As the temperature is increased, the free energy of rutile decreases faster than that of anatase, and rutile becomes the most stable bulk phase at temperatures above about 650~K. The effect of lattice vibrations at finite temperature is therefore crucial to the bulk phase stability of all the phases. 

The largest potential source of errors in our QMC calculations are the time step and finite size errors. We have demonstrated that these are well converged, and our results are not sensitive to them.
There are two sources of systematic error remaining in the zero-temperature QMC calculations that prevents them from being exact. First, our calculations are subject to the fermion sign problem and the fixed-node error that results from using FN-DMC to control it. Ideally the nodal structure of the trial wave functions would be fully optimized in QMC. This is becoming feasible for single Slater determinant wave functions in simple solids\cite{Devaux2015}, but more flexible many-body trial wave functions are also desirable. Today, we can not rigorously determine the relative nodal error between the TiO$_2$ phases, although we believe it to be small. Second, because we utilize pseudopotentials we must be concerned with the fundamental accuracy of the pseudopotential construction and the locality errors that result when evaluated in QMC. We expect considerable error cancellation because all of the polymorphs have, by definition, exactly the same composition. The locality error can be reduced by more accurate trial wave functions, and the atom-specific nature of the locality error suggests that this error may be targeted using methods inspired by quantum chemistry. Finally, the results for finite temperatures would presumably be more accurate if QMC methods could be used for the lattice dynamics; however, this is at the present not possible.
In the future, improvements to those remaining systematic error should be achievable when greater computational resources and improved methods are available. This would enable increased confidence in our results. These improvements are desirable for all QMC calculations and are not restricted to the current investigation.

In summary, our calculations indicate a potential resolution of the long-standing problem in reconciling DFT-based calculations with experimental observations. 
The fact that our QMC and many conventional density functional results obtain anatase as the most stable phase at 0~K suggests that these functionals might be qualitatively correct in giving the energetic ordering between anatase and rutile: it should not be automatically assumed that these calculations are in error, and we caution against using prediction of rutile as most stable at 0~K to validate new electronic structure methods. Our finite temperature calculations show that the lattice-dynamical contribution is important and should be taken into consideration when comparing with experimental data. We hope that these results stimulate further computational and experimental work on the phase diagram of this important material.

\section*{Acknowledgments}
An award of computer time was provided by the Innovative and Novel Computational Impact on Theory and Experiment (INCITE) program. This research has been funded in part and used resources of the Argonne Leadership Computing Facility, which is a DOE Office of Science User Facility supported under Contract DE-AC02-06CH11357. This research used resources of the Oak Ridge Leadership Computing Facility at the Oak Ridge National Laboratory, which is supported by the Office of Science of the U.S. Department of Energy under Contract No. DE-AC05-00OR22725. Sandia National Laboratories is a multiprogram laboratory managed and operated by Sandia Corporation, a wholly owned subsidiary of Lockheed Martin Corporation, for the U.S. Department of Energy's National Nuclear Security Administration under Contract No. DE-AC04-94AL85000. AB, LS, JK and PK are supported through Predictive Theory and Modeling for Materials and Chemical Science program by the U. S. Department of Energy Office of Science, Basic Energy Sciences (BES). OH was supported by the U.S. Department of Energy, Office of Science under Contract No. DE-AC02-06CH11357.

\end{document}